\documentclass[12pt, reqno]{amsart}
\usepackage{graphicx} 
\usepackage{epstopdf}
\usepackage{color}
\usepackage{tikz}
\usepackage{titlesec}
\usepackage{multicol}
\usepackage{url}
\usepackage{color}
\usepackage{hyperref}
\definecolor{refcolor}{rgb}{0.1,0.2,0.88}
\hypersetup{
    colorlinks,
    citecolor=refcolor,
    filecolor=refcolor,
    linkcolor=refcolor,
    urlcolor=refcolor
}
\usepackage[round]{natbib}

\definecolor{scolor}{rgb}{0.2,0.5,0.99}
\titleformat{\section}
{\color{scolor}\normalfont\Large\bfseries}
{\color{scolor}\thesection}{1em}{}
\titleformat{\subsection}
{\color{scolor}\normalfont\large\bfseries}
{\color{scolor}\thesubsection}{1em}{}

\definecolor{quotecolor}{rgb}{0.15,0.0,0.66}
\newcounter{questioncounter}

\definecolor{mytitlecolor}{rgb}{0.20,0.0,0.95}

\newcommand{\image}[3]{\begin{figure*}[ht]
\includegraphics[width=#2\textwidth]{#1}
\caption{\small{\label{#1}#3}}\end{figure*}}

\newcommand{\R}{\mathbb{R}}
\newcommand{\etc}{\textit{etc}}

\begin{document}




\begin{center}
\href{http://fqxi.org/community/essay/winners/2015.1#Stoica}{Third prize} in the FQXi's 2015 Essay Contest
\\\href{http://fqxi.org/community/forum/category/31424}{\textbf{`Trick or Truth: the Mysterious Connection Between Physics and Mathematics'}}
\\Chapter in \emph{Anthony Aguirre et al: ``Trick or Truth? The Mysterious Connection Between Physics and Mathematics''} at \emph{Springer, The Frontiers Collection}
\end{center}

\vspace{0.2in}

\begin{center}
\huge{\textbf{\color{scolor}{And the math will set you free}}}
\end{center}

\vspace{0.1in}

\begin{center}
\textit{Cristi \ Stoica}
\\
{\footnotesize Department of Theoretical Physics, National Institute of Physics and Nuclear Engineering -- Horia Hulubei, Bucharest, Romania}
\\\footnotesize{\href{mailto:holotronix@gmail.com}{h\,o\,l\,o\,t\,r\,o\,n\,i\,x\,@\,g\,m\,a\,i\,l\,.\,c\,o\,m}}
\end{center}




\vspace{0.2in}

\section{Introduction}


How mathematical is the physical world?
Is mathematics essential to describe the universe, or it is just a tool which is not even necessary? There is no unanimously agreed answer to this question. The spectrum of opinions ranges from ``mathematics is just a tool helping us to classify our observations'' to ``the world is nothing but a mathematical structure''.
Like for any debate, part of the tension between different views is the implicit usage of different definitions. So, to avoid possible confusion, we should define both mathematics and physics.


Is mathematics discovered, or invented?
Are mathematical structures entities which have their own existence, eternal and unchanging? Then, how can we know them? If we can access them with our thoughts, shouldn't they then be connected to our minds somehow? Can we discover and explore them? Or it is us who invent them? Or maybe we invent the axioms, and then we discover the consequences? Let's try to find out.

\section{A universe in a dot}

Think at a number. Did that number exist before you picked it, or it is you who invented it? Don't rush with the answer, because it is not as easy as it might seem.

Consider the most brilliant text ever written. Is it the Bible, or the complete works of Shakespeare, or perhaps the complete collection of arXiv articles? Any of these can be written in a computer. The computer uses a code to assign numbers to letters, for example $A\to 65$, $B\to66$, ..., $Z\to 90$, $a\to 97$, ..., $z\to 122$. Internally, it represents these numbers in binary form, as strings of $0$ and $1$, so that for example $A$ becomes $01000001$, $B$ becomes $01000010$ \etc. Let's take any text and assign to it a number between $0$ and $1$, of the form $0.d_1d_2\ldots$, where $d_1d_2\ldots$ is the decimal representation of the binary representation of the text.

So, for example, the King James Bible starts with
\begin{quote}
1. In the beginning God created the heaven and the earth.
\end{quote}
Therefore, its number is 
$$0.192110078742304518747878005390999191523668554050387237...,$$
which obviously is between $0$ and $1$ (see fig. \ref{bible.png}).

\image{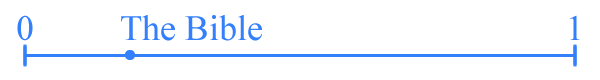}{0.7}{The Bible as a number between $0$ and $1$.}

Not only all literary creations are there, but also all musical works, all movies, anything. 

Even this essay is there. So should you stop reading right now, and just look at that line?

That line contains a complete description of your entire life, your past and future, and the moment of your death, since all written information that survives us is nothing but a text but a number but a point.

Imagine now a distant future of mankind, when we know as much as possible about the universe. Suppose we will collect all information we will have about the universe, and also all human creations, recorded in any way, and store everything on a digital support. It's associated number will also contain everything, including any piece of literature, art or science ever written. A universe in a dot (fig. \ref{everything.png}).

\image{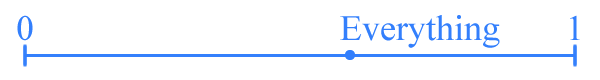}{0.7}{All human creation and knowledge is a point on this line segment.}

By not allowing you to do anything you want, mathematics feels often like a prison. But you have absolute freedom even in a small segment like $[0,1]$!

\section{So, is mathematics discovered, or invented?}

Do the numbers between $0$ and $1$ already exist, or we keep inventing them? If they preexist, then we have to admit that any poem is already there, among the numbers between $0$ and $1$. If we invent them, then we have another problem. Given that we label the moments of time with numbers, did we invent every moment from the birth of the universe until now? Since a position in space is given by its coordinates, numbers as well, did we invent every point in space too?

It seems that one cannot simply say that mathematics is invented, since we can't go outside its boundaries, no matter how creative we are. We can't say we discover it either, maybe that we construct it. Or even better, we reconstruct it. It is as if we dance freely, only to find out that we stepped on footprints that were already there. Mathematics is already there, eternal and unchanging. What we invent is the discovery of mathematics.

\section{But what is mathematics?}

I will take here the position that {\em mathematics is the science of mathematical structures}. A {\em mathematical structure} is a set $S$, together with a collection of relations between the elements of $S$. A {\em relation} is a collection of ordered sets of $n$ elements of $S$ \citep{birkhoff1946universalalgebra}.

As simple and restrictive as this definition may seem, all the mathematics used by physicists is contained in it. To show how this may be would take many books, but there is room for a few simple examples. The real numbers form a set $\R$. There is a relation of order, which says that $3$ is smaller than $4$ for example. The relation of order can be seen equivalently as the collection of all pairs of the form $(a,b)$, where $a<b$. How about mathematical operations? Addition is just the collection of triplets $(a,b,c)$, where $a+b=c$. Similarly for multiplication. So, the real numbers form a mathematical structure in the sense defined above. Although I will not detail here how, the entire mathematics currently used in physics is like this, including groups, vector spaces, Hilbert spaces, spacetime manifolds, fiber bundles, and so on.

\section{What is physics?}

Very roughly speaking, physics should be a collection of true statements about the physical world. It should account for the observations and experimental findings we made so far, but also for those to be made. While we can collect the results of our observations in catalogs, this is not enough to predict the results of future ones.

For example, Tycho Brahe collected the positions of planets and stars from very accurate observations spanned along many years. But to predict the future positions of the planets we had to wait Kepler to discover the rules governing their movement. Unsurprisingly, these were mathematical rules. Due to the work of Newton, the mathematical formulae of Kepler were obtained as consequences of simpler and much more general laws.

Hence, mathematics entered initially in physics as a way to record the observations in a precise way. Then, it was used as an ordering principle, which organizes the quantitative observations. But once they are organized, they appear as consequences of deeper laws which are mathematical in nature. This allows not only to organize the previously known data, but also to make predictions.
All physical laws we discovered so far follow the same pattern.

This is why mathematics is so intimately tied with physics.

\section{Is mathematics merely a tool in physics?}

As beings immersed in the universe and observing it, we experience the fact that there are regularities. We learn how to use these regularities to anticipate and control the flow of events. Luckily we don't need them to be truly accurate for most of our daily objectives; it is enough if they are practical.

From this perspective, mathematics seems as a useful tool in organizing our knowledge and giving more or less accurate quantitative predictions. We owe mathematics the freedom brought to us by the scientific and technological progress. But the fact that mathematics is useful doesn't mean the universe is mathematical, not even partially mathematical.

Maybe these useful mathematical models of various phenomena are only approximations of a deeper description. And indeed, when you try to use mathematics to simulate phenomena, you know you use discrete structures to simulate objects that appear to be continuous, you know that the simulation contains a smaller number of degrees of freedom, that it had to ignore some parameters and so on. Yet, the simulation is useful, even though we know it is only an approximation.

But why couldn't that reality we approximate by mathematics be itself mathematical?

\section{Is there something that can't be described by mathematics?}

The idea that mathematics is merely a tool, which is replaceable and is not even the best one, found supporters among philosophers and even physicists.
Recently, Smolin and Unger \citep{Smolin2015FQXi,Unger2015FQXi,UngerSmolin2014SingularUniverse} brought some strong arguments supporting the idea that mathematics not only is not discovered, but it is not even able to describe the universe in a unified way.

Smolin said \citep{Smolin2015FQXi}:

\begin{quote}
there is no mathematical object which is isomorphic to the universe as a whole, and hence no perfect correspondence between nature and mathematics.
\end{quote}

\begin{quote}
There are no eternal laws; laws are subsidiary to time and to a fundamental activity of causation and may evolve.
\end{quote}

Scientists found regularities in the way nature works, regularities successfully tested repeatedly, having predictive power. Mathematics always played a major role in this. We know laws that were never contradicted by our observations, and their validity seems to be eternal, but this doesn't mean they will not be contradicted by experiment someday, or that there are no corners of the universe where they don't apply. So there is no way to disprove Smolin's statements, but there's no way to prove them either. Mathematics instead proved its predictive power together with any successful theory in physics.

Our experience shows that when a law of physics is broken in a particular regime, hence turns out to be approximate and not universal, it is to make room for a more general law, which explains more. For example, the predictions of general relativity went beyond the limits of Newtonian gravity. The laws based on a flat space were replaced with more general laws, on curved spacetime, and the very curvature explained gravity as a form of inertia. When our understanding of general relativity turned out to be limited due to the occurrence of singularities, mathematics provided a more general geometry, one which works at singularities too \citep{Sto11a,Sto12b,Sto13a}, and this allowed us to better understand the big-bang singularity \citep{Sto11h,Sto12a}, and the black hole singularities \citep{Sto11e,Sto11f,Sto12e,Sto14a}, and to open new avenues to quantum gravity \citep{Sto12d}.

One central argument invoked by Unger and Smolin is that if mathematical structures would have existence, they would be independent of the physical world. Therefore, any claim that we can describe the universe mathematically would rely on the belief in some mystical powers which would allow us to access worlds outside the universe. The problem with this argument is that people supporting the strong connection between physics and mathematics actually don't think that mathematical and physical worlds are separated, and some claim they are even the same \citep{tegmark2008mathematical}. So, the argument invoked by Smolin and Unger doesn't contradict the hypothesis that the universe is (at least isomorphic to) a mathematical structure.

According to Unger, there are things that can't be described by mathematics, such as {\em time} and {\em particularities} \citep{Unger2015FQXi}:

\begin{quote}
Mathematics is an understanding of nature emptying it out of all particularity and
temporality: a view of nature without either individual phenomena or time.
\end{quote}

But can we find a property of {\em temporality} that can't possibly be described by mathematics? In fact, time was best understood due to mathematics, in thermodynamics, relativity, and quantum mechanics. Any phenomenon related to time, for example memory, can in principle be described mathematically. Probably what seems to escape any mathematical description is {\em our feeling} that time flows. But this is another issue, which is not confined to the feeling of time flow only: the {\em hard problem of consciousness} \citep{Chalmers1995HardProblem}. However, any feeling we may have, there are neural correlates associated to it, and hence, physical correlates. And these physical correlates are in the domain of known physics, which is includes time and is strongly mathematized. But by this I don't claim we can explain consciousness, with or without mathematics. My only claim is that its physical manifestations are describable by mathematics, at least in principle.

Unger writes \citep{Unger2015FQXi}

\begin{quote}
Our mathematical and logical reasoning has a characteristic that places it in sharp
contrast to our causal explanations. A cause comes before an effect. Causal explanations make
no sense outside time; causal connections can exist only in time.
\end{quote}

Actually, there is no conflict between mathematics and causal explanations. The equations describing the time evolution of any physical system reveal the causal connection between the initial conditions and any future state of a system. If causality is still not clear here, consider general relativity, where the very geometry of spacetime gives the causal connections. Of course, the metric tells which events are causally connected, but it doesn't tell the {\em arrow of time}. The reason why time has a privileged direction is not yet understood, but statistical mechanics explains it by the fact that at a time the entropy was very low. Penrose tried to explain this in a geometric way, and conjectured the {\em Weyl curvature hypothesis} \citep{Pen79}, as accompanying the big-bang singularity. A version of this hypothesis was shown indeed to be true for a large class of big-bang singularities, which are not necessarily homogeneous or isotropic \citep{Sto12c}.

The other thing which was supposed to be impossible to describe by mathematics are the {\em particularities}. The reductionist approach aims to explain the diversity of nature in terms of a few fundamental universal laws and building blocks. How is it possible that universal and timeless laws explain particular manifestations which appear, evolve, and vanish? Well, it is not about explaining, unless we take the word ``explanation'' in its weak sense of ``description''. Mathematics allows us to describe everything about these particular instances in a reductionist way. But we don't know yet well enough how to obtain from the equations describing how matter evolves in time the various particular forms. Emergent phenomena are still not well understood. But the little we understood so far was done with the help of mathematics, and it doesn't contradict it at all \citep{Bar2004EmergenceMathematical,Cucker2007EmergenceMathematics,Ellis2012Emergence}. I don't know of a non-mathematical alternative explanation of emergence. At least mathematics can be used to describe the particular forms, even if it can't tell yet how and why they appear. Postulating emergence or particular forms as fundamental is not a solution, we need to explain as much as we can before doing this. Some think even the hard problem of consciousness can be solved in a reductionist way \citep{Dennett2001Consciousness,Dennett2013IntuitionPumps}, although so far only {\em easy problems} could be explained. But no matter how scared we may be that mathematics of physical laws is too rigid to allow consciousness, at least we know that there is room for free will, whatever this may be \citep{Hoefer2002Freedom,Sto08f,Sto12QMa,Aaronson2013ghostQM}.

\section{Is everything isomorphic to a mathematical structure?}

From the definition of mathematics I am using here it's easy to see that everything behaves identically to a mathematical structure, everything is isomorphic to a mathematical structure.
To see this, consider that we make a complete list of truths about certain domain, which may be even the entire universe. It should be no contradiction between the propositions in the list. Then, there is a mathematical structure for which the same propositions are true.
We call the list of propositions a {\em theory}, and a mathematical structure for which these propositions are true, a {\em mathematical model} of that theory -- in the sense of {\em model theory} \citep{chang1990model}.

Some may think that there are things in the universe which can't be described by mathematics. But can you name those things? To name them, you would have to provide a list of their properties, of propositions which hold for them. If the universe is describable by a list of propositions, then there is a mathematical structure describable by the same propositions.
But then, couldn't we find something to say which is true about our universe, but not about that mathematical structure? The answer is no. Even if we manage to extend the list with new truths about the universe, there is a mathematical structure which is isomorphic to the universe described by the extended list of propositions \citep{Sto13b,Tegmark2014OurMathematicalUniverse}.

I think at this point many of the readers feel that I am too reductionist, and that there are so many things that can't admit mathematical descriptions. Arts, poetry, love, music, intuition, faith, how can all these be completely mathematical? Well, I don't mean that we have a mathematical description of them. But think at an emotion related to music, poetry, or love. If we describe it completely by a list of true propositions, then you can say it is mathematics. If we can't, it's only because of practical limitations. We know what a feeling is: some chemistry of the brain. But it would be impractical to search for the complete description of the conditions that boost your brain's levels of dopamine, serotonin, oxytocin, and endorphin. However, this doesn't mean that there is no such a description. And if there is, there is also a mathematical model of it, waiting for us.

So, the idea that everything is isomorphic to a mathematical structure may not sound that crazy. Of course, we don't know yet that structure, and it may be too complex to be possibly known, but this doesn't change the things.

If you feel uncomfortable to work with a mathematical structure isomorphic to a phenomenon or a part of the world, you can work safely with the list of propositions describing it, it is the same thing. Sometimes you may prefer to use Tycho Brahe's tables, Kepler's laws, or Newton's laws, depending on the level of abstraction of the problem you want to solve.

\section{Can the universe be mathematical?}

Even if we admit the idea that our universe is isomorphic to a mathematical structure, does this mean that it is mathematical? Maybe the universe is something that is just describable by a mathematical structure, without being such a structure. But if there is a complete isomorphism between the universe and a mathematical structure (unknown to us, at least so far), there is nothing to tell the difference between them. And since there is no difference that can be tested by experiments, one should admit that the universe is mathematical (I invoke Leinbiz's {\em identity of indiscernibles} principle). If it looks like a duck, swims like a duck, and quacks like a duck, then it is a duck, isn't it? Whenever we think that there is something more than this, either it has some effects and relations with the rest, which can in principle described by a list of propositions, or it has no effects whatsoever. And if it doesn't have effects, if we limit ourselves to the physical world, we can safely ignore it. So, for now on, when I will say that the universe is mathematical, it will be in this sense.

\section{Does the hypothesis of a mathematical universe make predictions?}

A {\em principle} is a general rule which describes in a concise way the data collected in repeated experiments and observations of a class of phenomenona. The principle is a hypothesis, but it can be tested. By using logic, one can derive consequences from the principles, and make predictions of the results of other experiments. If the predictions are not confirmed, we have to reject the principle which led to them.

But is the hypothesis that the universe is mathematical a testable one?

We have seen that any kind of world, as long as it is free of contradictions, is isomorphic to a mathematical structure. This means that this hypothesis is a plain truth that doesn't make predictions at all, and doesn't explain anything. But this doesn't mean it is useless, rather it means that it is the foundation, the framework of any possible theory in physics.

\section{Is there a theory of everything?}

We know for many decades that our universe is very well described by two theories, general relativity and quantum theory. Each of them has its domain of applicability and validity. But shouldn't be only one theory governing the laws of our universe? It seems natural that this should be true, and that unified theory is often called ``the theory of everything'', and is supposed to include general relativity and quantum theory and everything else.

Because we couldn't find so far this theory, there are physicists considering that it doesn't exist, and maybe the universe obeys two or even more sets of laws. This doesn't make much sense, since if the universe obeys two or even more independent sets of laws, there must be two or more disconnected mathematical structures modeling them. But we can't live simultaneously in two disconnected worlds.
Hence, there must exist a mathematical structure which satisfies our observations about both the quantum world, and the general relativistic one. Maybe these theories are somehow limits of this theory. But the unified theory must exist, even if we don't have it yet.

\section{What can stop us from finding the theory of everything?}

The standard way of science is this: we look at the data of our experiments and observations, we guess the rule, we derive other consequences of that rule, we design and perform experiments to test those consequences, and if the predictions are invalidated by experiments, we reject the rule and try another one, and so on. But this can't ensure that we will find the ultimate theory of everything (TOE). There is no guarantee that we can test all the truths about the universe. It is clear that, for example, we can't check all the details of events that happened long time ago, but I am not talking here about particular configurations of the universe. 
What I mean is that it is not sure that we can test even the universal laws. Being able to guess them and then test them would mean either that we are that lucky, or that the universe wants to be completely understood by us, who are just tiny waves on its surface.

However, it may be possible that we will be able to find the fundamental physical laws. The reason is that the universe seems to be very regular. The physical laws appear to be the same at any point and at any time. They appear to be the same for observers moving at different velocities, even though the movement is accelerated. This is the principle of relativity, which is probably the most well tested principle of physics. The laws are the same everywhere, anytime, and for anyone. This simplifies very much the process of testing the other physical laws.

Considering the diversity of phenomena, the number of independent laws which explain almost anything is relatively small. Moreover, each progress that we made unified concepts and replaced the known laws with fewer, more general ones. For example, special relativity unified time and space, energy and momentum, and the electric and magnetic forces. General relativity unified gravity and inertia. Quantum theory unified waves and particles, frequency and energy, explained the atomic spectra. All of the forces in nature, no matter how different they may appear, reduce to gravity (which is spacetime curvature), and three gauge forces. Eventually, all of our knowledge about the physical laws is encoded in quantum physics, the spacetime curvature, and the standard model of particle physics, which involve a much smaller number of concepts than the entire field of physics.

This economy of laws is possible only because of mathematics. When we learn physics, and see how various concepts reduce to more fundamental ones, the general pattern is not only that more complex objects are composed of atoms, which are made of elementary particles. The general pattern is that a handful of laws combine mathematically and give the huge diversity of observed physical phenomena. The best and only way to understand them is with the help of mathematics. Some laws can be explained to non-experts using only words, but with the equations, the things are much clearer, more precise, and avoid naive misunderstandings.

So, if we can understand the universe, it is because this immense complexity can be reduced to a small number of laws. And this reduction is made possible by mathematics.

Some may now say that we will never be able to find the unified theory, because of {\em G\"odel's incompleteness theorem} \citep{godel1931}. The argument is that, since by this theorem any mathematical theory which contains arithmetics is incomplete, this means that we will never be able to find a complete theory of the universe. Well, this is actually a misunderstanding of the famous theorem. Remember that at the end of the XIXth century, it was believed that everything is explained by classical mechanics and electrodynamics. Suppose that there were no relativistic and quantum effects, and the world was really classical. Then, what we had at that time would be the unified theory. G\"odel's theorem couldn't prevent this. Did something change after we realized that there are relativistic and quantum effects? Why the G\"odel incompleteness theorem would suddenly became relevant? It is true that the unified theory, whichever will be, is likely to contain arithmetic, and hence be incomplete in G\"odel's sense, but this would not make it a less unified theory. G\"odel's theorem simply states that no finite system of axioms allows any truth valid in that system to be found by finite length proofs.

Hawking gave the following argument against a theory of everything, based on G\"odel's theorem \citep{Hawking2002Godel}:

\begin{quote}
we and our models, are both part of the universe we are describing. Thus a physical theory, is self referencing, like in G\"odel's theorem. One might therefore expect it to be either inconsistent, or incomplete. The theories we have so far, are both inconsistent, and incomplete. 
\end{quote}

However, we are just looking for a theory describing the general laws, and not a complete description of this particular instance of the universe, which includes what every human thinks about the universe and themselves. This would not be feasible anyway for practical reasons. A TOE should not be expected to contain the complete description of the state of the universe, only the universal laws, and perhaps various approximate models of particular configurations, for example cosmological models, models of stars and galaxies {\em etc}. This is consistent, for the same reason why it is consistent for a computer to contain in its memory the complete technical specifications of itself and its operating system, and an approximate representation of its state.
Hence, it is not clear how the fact that we and our theories are part of the world we are trying to describe can lead to the situation in the proof of G\"odel's theorem.

\section{Tegmark's mathematical universe hypothesis}

The idea that the universe is nothing but a mathematical structure leads to many difficult and interesting questions. For example, why this particular mathematical structure and not another? Tegmark \citep{tegmark1998TOEensemble,tegmark2008mathematical,Tegmark2014OurMathematicalUniverse} proposes a very democratic point of view, that all possible mathematical structures exist physically too \footnote{This is a mathematical expression of David Lewis's {\em modal realism} \citep{lewis2008convention,lewis2013counterfactuals,lewis1986plurality}.}. That is, {\em mathematical existence} (which means just {\em logical consistency} -- the absence of internal contradictions) equals {\em physical existence}. Now you may think that to admit that all possible mathematical structures exist is too much of a waste. But the idea is in fact very economical, if you think that to choose a single mathematical structure out of an infinity of them requires to specify its definition, or its axioms, while to allow all of them to exist doesn't require to specify all of them, and in fact doesn't require any information. What's easier, to specify the complete works of Shakespeare, the Bible, and any other piece of art, literature, and science, or to specify the interval $[0,1]$ from fig. \ref{everything.png}?

This makes us wonder where are the other mathematical structures. But the answer is simple: ``out there'', and if we can't see them, it's because our universe is disconnected from the other possible universes. But if the other structures are disconnected and we can't reach out to check them, then this idea can't be tested.

Tegmark has an interesting idea of a proof of his mathematical universe hypothesis, which in the same time aims to answer the question ``why, among all possible worlds, we live in this particular world?''. Tegmark's answer is an anthropic one: we live in this world because it is favorable to intelligent observers ({\em self-aware substructures}, or SAS \citep{tegmark1998TOEensemble}). For example, the planetary orbits would be unstable if space would have a different number of dimensions \citep{tegmark1997DimensionalityOfSpacetime}.

But why the equation of Laplace (which is responsible for the inverse-square law in three dimensions, and hence for the planetary orbits) has to be true, and intelligent beings have to inhabit planets orbiting around stars?
Given that we already accepted all possible mathematical structures as physical universes, maybe in the vast majority of the universes the equation of Laplace doesn't hold or is not applicable.

Such arguments that most possible worlds supporting SAS could not be so different than ours rely on varying one constant while keeping the others fixed, and then showing that the optimal value for that constant to allow SAS is the one it has in our universe. 
But if we vary the entire structure, we can find any kind of worlds which would support one form of intelligence or another.

If we are Turing machines, we could exist in any universe in which such Turing machines can exist. And so many mathematical structures are able to contain simulations of any kind of Turing machines. For example, the {\em Turing complete} ones can simulate any other Turing machine and any algorithm. Simple cellular automata, governed by very simple rules, like Conway's {\em game of life} \citep{conway1970life} and {\em Rule 110} \citep{cook2004universalityRule110,cook2009concreteRule110}, are Turing complete. This means that the only prediction made by the mathematical universe hypothesis is that our universe has to be Turing complete, or at least complex enough to simulate intelligences like us. Why would be a universe like ours better at simulating intelligent observers, than the Rule 110 cellular automaton, or than Conway's life game? There is no reason, since they are computationally equivalent. If our intelligence is nothing but an expression of us being Turing machines, we could exist in any cellular automaton which is Turing complete, including in the Rule 110 cellular automaton.

Hence, to prove the mathematical universe hypothesis using Tegmark's idea it is not enough to consider only small variations of our universe, but to take into account many other possibilities to sustain intelligent observers.

\section{Counting minds}

A possibility to modify Tegmark's method of verifying the mathematical universe hypothesis is the following.

Consider the mathematical structures satisfying the following conditions
\begin{enumerate}
	\item 
they can be interpreted as Turing machines,
	\item 
they admit configurations which contain self-aware substructures (which in turn can be interpreted as Turing machines).
\end{enumerate}

The notion of {\em self-aware Turing machines} deserves a rigorous definition, but I don't think we have one at this time. However, they have to be able to gain a significant understanding \footnote{Whatever ``understanding'' may mean for Turing machines, perhaps ``to understand'' means ``to simulate''.} of themselves and of their universe, and possibly other universes. So I suggest that a pretty good approximation is to consider them to be Turing complete. In this case, any such SAS would be capable of understanding (that is, of simulating) any other SAS. This means that all mathematical structures under consideration are computationally equivalent, so how can we then distinguish them and tell that we live in the one which is most likely to support intelligent observers?

I think the solution is to find a way to count all the possible configurations these mathematical structures can have, and to count among them those containing self-aware observers \footnote{Don't forget to take into consideration the possibility of {\em Boltzmann brains} -- intelligent entities which appeared spontaneously, without a history, because of fluctuations \citep{albrecht2004BoltzmannBrain}.}. For example, among all possible configurations of a certain mathematical structure, find what is the percentage of those containing SAS. Then, given all self-aware substructures, we find in which particular mathematical structures it is more probable for them to exist. If Tegmark is right, the mathematical structures we will find will be very similar to our universe.

The problem is that we don't know how to count the probabilities in the infinite collection of all possible configurations of all possible mathematical structures. For example, if the number of configurations for two structures is infinitely countable, then for any configuration of the first structure there is one for the second structure and the other way around. To find such a correspondence, we label each configuration of the first structure uniquely with the numbers $1,2,3,...$, and similarly to the second structure. Then, we put into a one-to-one correspondence the configurations of the two structures having the same label, using the function $f(n)=n$. But we could as well use the function $g(n)=2n$, and in this case, the second structure may appear to have twice as many configurations as the first structure. To calculate probabilities in infinite sets you need a {\em measure}, which we don't know how to define in a correct way in this case.

Suppose that we have found indeed that the typical SAS live in mathematical structures similar to our universe. In this case, if another mathematical structure contains SAS, the cheapest way to contain them is by simulating those mathematical structures similar to our universe. But a mathematical structure can exist in two ways: by itself, or as a substructure of another structure. And to each particular configuration of a mathematical structure which exist by itself, there correspond an infinite ways in which it can exist as a substructure of other structures. For example, if there is one Euclidean plane which exists by itself, there is an uncountably infinite number of Euclidean planes which exist as subspaces of an Euclidean space. And the Euclidean space in its turn is more likely to exist as a subspace of a hyperspace and so on. Therefore, it will be infinitely more likely that SAS exist in mathematical structures which contain simulations of structures like our universe, rather than in structures like our universe themselves.

This only shows that there are important difficulties in trying to prove Tegmark's mathematical universe hypothesis, but not that it is impossible to be proven.

\section{Concluding remarks}

The history of physics shows that as theories evolve, they become more and more mathematized. Mathematizing our theories increases their predictive power, their rigor, their applicability, and unifies various concepts, making them more economical and simpler (but more abstract). Whether this is because mathematics is very versatile, or because physics has a prominent mathematical character, or even because the universe is a mathematical structure, is an open debate.

But I would like to suggest that, just like in our understanding of nature supernatural explanations were gradually replaced by natural ones, in physics, ``supermathematical'' descriptions are gradually replaced by mathematical ones.
I think that we should admit ``supermathematical'' descriptions as final only if we are sure that we exhausted any hope for a mathematical description. And this may never happen.

\vspace{0.2in}
\textbf{\color{scolor}{Acknowledgments.}}
I wish to thank Alma Ionescu for helpful comments.

\newpage
\scriptsize
\begin{multicols}{2}

\end{multicols}

\newpage
\begingroup
\def\enotesize{\normalsize}
\endgroup

\end{document}